\documentclass[twocolumn,preprintnumbers,
endnote,nofootinbib,prl]{revtex4}
\usepackage{graphicx}
\usepackage{color}

\newcommand{\lsim}{\lesssim}
\newcommand{\gsim}{\gtrsim}

\newcommand{\eq}[1]{Eq.~(\ref{#1})}

\newcommand{\ord}[1]{\mathcal{O}{(#1)}}
\newcommand{\beq}{\begin{equation}}
\newcommand{\eeq}{\end{equation}}
\newcommand{\bea}{\begin{eqnarray}}
\newcommand{\eea}{\end{eqnarray}}

\newcommand{\mpbh}{M_{\rm PBH}}
\newcommand{\mns}{M_{\rm NS}}
\newcommand{\mabh}{M_{\rm ABH}}

\newcommand{\msol}{M_\odot}
\newcommand{\mtot}{M_{\rm tot}}
\newcommand{\rsch}{R_{\rm Sch}}

\newcommand{\kpc}{~\text{kpc}}

\begin{document}

\pagestyle{plain}

\title{\boldmath Gravitational Waves from Primordial Black Holes and New Weak Scale Phenomena}

\author{Hooman Davoudiasl
\footnote{email: hooman@bnl.gov}
}

\author{Pier Paolo Giardino
\footnote{email: pgiardino@bnl.gov}
}

\affiliation{Department of Physics, Brookhaven National Laboratory,
Upton, NY 11973, USA}


\begin{abstract}

We entertain the possibility that primordial
black holes of mass $\sim (10^{26}$--$10^{29})$~g, with Schwarzschild radii 
of $\ord{\text{cm}}$, constitute $\sim 10\%$ or more of cosmic dark matter, as allowed by various constraints.  These black holes would typically originate from cosmological eras corresponding to temperatures $\ord{10-100}$~GeV, and may be associated with first order phase transitions in the visible or hidden sectors.  In case these small primordial black holes get captured in orbits around neutron stars or astrophysical black holes in our galactic neighborhood, 
gravitational waves from the resulting ``David and Goliath (D\&G)" 
binaries could be detectable at Advanced LIGO or Advanced Virgo for hours or more, possibly over distances of $\ord{10}$~Mpc encompassing the Local Supercluster of galaxies. The proposed Einstein Telescope would further expand the reach for these signals.  A positive signal could be further 
corroborated by the discovery of new particles in the $\ord{10-100}$~GeV mass range, and potentially also the detection of long wavelength gravitational waves originating from the first order phase transition era.

\end{abstract} \maketitle

The presence of cosmic dark matter (DM) is firmly established by various cosmological and astronomical 
observations \cite{PDG}.  However, all existing evidence for DM is from its gravitational effects.    
While it is widely believed that DM has non-gravitational interactions that governed its production in the early Universe, 
all attempts to uncover those interactions have been unsuccessful.  This situation motivates one to entertain 
the possibility that DM is of a purely gravitational nature.  In particular, if DM is composed of primordial black holes (PBHs) \cite{Carr:1974nx,Meszaros:1974tb,Carr:1975qj,Khlopov:2008qy}, formed via 
gravitational collapse of primordial matter around over density perturbations in the early Universe, it may only manifest itself through its gravitational effects.    

The above PBH scenario removes the need to postulate new particles and interactions associated with DM, which is often invoked as strong motivation to search for physics beyond the Standard Model.  This intriguing possibility is quite constrained by various observations \cite{CPT,Griest:2013esa,Carr:2016drx,Subaru-pbh} over most of the viable parameter space.  However, some parts of the parameter space allow for PBHs to be a significant component of DM.  In fact, allowing for deviations from a monochromatic spectrum, which is expected to be the case in realistic scenarios \cite{Carr:2016drx}, some narrow ranges of parameters could possibly allow for all DM to be composed of PBHs. 

The primordial nature of the DM black holes implies an interesting correspondence between 
the masses of PBHs and the era in which they were produced.  That is, since PBHs are assumed to be formed by the collapse of matter and energy over a Hubble volume, the 
mass $\mpbh$ of a PBH 
is a measure of the horizon size, and hence the temperature of the Universe at the time of the PBH formation.  

PBH masses that could potentially originate from first order phase transitions \cite{PT-PBH} at temperatures $T\sim \ord{10-100}$~GeV could offer an interesting window into experimentally accessible particle physics.  This range of $T$ can be associated with extensions of the electroweak sector and the Higgs potential and may also lead to long wavelength primordial gravitational waves, which may been within the reach of future space-based observatories \cite{spaceGW}.  Those extensions may play a role in electroweak baryogenesis and also provide new possibilities for microscopic DM candidates, with PBHs comprising a sub-dominant, but potentially significant, DM population.  The  above range of $T$ roughly corresponds to \cite{PT-PBH}   
\beq
10^{26}~\text{g}\lsim \mpbh \lsim 10^{29}~\text{g}\,.
\label{Mpbh}
\eeq
Current bounds, including the recent micro-lensing searches 
from observations of the Andromeda galaxy 
by the Subaru Hyper Suprime-Cam \cite{Subaru-pbh}, still allow about $5-10\%$ of DM to be comprised of PBHs, for masses in the above range.  With the assumption of a distribution for $\mpbh$, it might be possible that a somewhat larger fraction of DM is made up of PBHs over the range (\ref{Mpbh}).  
The Schwarzschild radius $\rsch$ of a black hole 
scales linearly with its mass $M_{\rm BH}$ and 
in the range (\ref{Mpbh}) above, 
the corresponding Schwarzschild radii are $\rsch\sim 0.01-10$~cm.   

In this work, we will consider values of $\mpbh$ in the range (\ref{Mpbh}) and explore the possibility that a neutron star (NS) or an astrophysical black hole (ABH) in our galactic neighborhood 
may have captured a PBH of such masses in an orbit around them.  As we will show, 
the gravitational wave signals from these ``David \& Goliath (D\&G)"\footnote{Note that, in our version of the confrontation, Goliath fares far better than in the original story.} 
binary systems can be detectable at Advanced LIGO (aLIGO) or Advanced Virgo (AdV), at their design sensitivity up to distances of $\sim 10$~Mpc, covering the Local Supercluster of galaxies.  The envisioned future Einstein Telescope 
could possibly extend the reach for the parameters considered here to $\ord{50}$~Mpc, going beyond the Local Supercluster.  

We point out that there can be two possible formation mechanisms 
for D\&G binaries: (a) through radiative capture; see for example Refs.~\cite{MTurner1977,O'Leary:2008xt,Cholis:2016kqi} and (b) through adiabatic contraction and dynamical friction.  The first possibility has been examined extensively in the context of solar mass black holes and could in principle be applicable here.  We suggest the second possibility based on proposed constraints for PBHs, where one estimates the likelihood that a PBH be captured during star formation and later end up within the compact remnant, such as a white dwarf, destroying it \cite{CPT}.  It seems plausible that one may also use this process to form D\&G binaries that will later coalesce and yield our signal.  We will focus on the first mechanism (a), however possibility (b) may also result in viable candidates.  Hence, our estimate 
for the rate of D\&G inspiral events could be considered conservative in this sense.  Without a more dedicated analysis - which is outside the scope of this work - it may not be possible to determine which of the (a) or (b) options yield the dominant rate and what a reliable estimate of that rate would be.

As we will discuss in the appendix, formation of D\&G binaries via radiative capture is likely a rare occurrence and our estimated rate might be $\sim 10^{-4}$ per year or less. However, our proposed signals could be detected using the existing aLIGO/AdV facilities and do not require dedicated new experiments. In light of the above, and given the major impact of a potential PBH discovery on our understanding of the Universe, an examination of our proposal appears worth while, even if PBHs constitute only a subdominant contribution to DM.\footnote{See Refs.~\cite{Bird:2016dcv, Clesse:2016vqa} for recent works that examine whether the observation of gravitational waves \cite{Abbott:2016blz} from the merger of black holes with $\sim 30$ solar masses  
corresponds to detection of DM composed of PBHs. Ref.~\cite{Nakamura:1997sm} examined  solar mass PBH binaries and Ref.~\cite{Inoue:2003di} considered sub-lunar mass PBH binaries.  For other possible signals of PBHs see Refs.~\cite{Kesden:2011ij,Belotsky:2014kca}.}

Let us begin with some general information about the astrophysical objects of interest.  
The known NS and ABH populations have masses $\mns \sim 1-2 \msol$ and $\mabh \gsim 10 \msol$ respectively, 
where $M_\odot \approx 2\times 10^{33}$~g is the solar mass.  For concreteness, in what follows we will choose 
\beq
\mns = 1.5 \msol \quad \text{and} \quad \mabh = 10 \msol\,,
\label{masses}
\eeq
as our reference values, however recent gravitational wave observations \cite{Abbott:2016blz} suggest that values of 
$\mabh \sim 30\msol$ are not necessarily uncommon.  We note that the nearest known NS and ABH are at distances $d_{\rm NS}\sim 0.3$~kpc and $d_{\rm ABH}\sim 1$~kpc, respectively. 
These objects are known due to optical observations.  In principle, there could be isolated compact stellar objects that do not 
emit detectable optical signals and may be closer to the Solar System.  In any event, we will use 
\beq
d \gsim 5 \kpc
\label{d}
\eeq
as a reasonable lower bound on the distance to potential binaries of interest in  our work.

We are interested in signals from an 
Extreme Mass Ratio Inspiral \cite{Hughes:2000ssa}.  Here we comment on a possible 
formation mechanism for such a binary, option (a) mentioned before, 
by emission of gravitational radiation during the 
initial D\&G encounter \cite{MTurner1977,O'Leary:2008xt,Cholis:2016kqi}.  One finds that 
the resulting binary orbits initially have $\ord{1}$ eccentricities $e$.  The orbits get circularized as the binary evolves, 
however for very hierarchic mass ratios the rate at which the eccentricity decreases 
$de/dt \propto -\mpbh/\mabh$ \cite{Peters:1964zz} is slow and the eccentricity may still be sizable at the final merger.   Hence, the circular orbit 
approximation may not be very accurate for the systems we focus on.  One of the main consequences of having $e\sim 1$ 
is that the gravitational radiation emitted by the binary is not dominated by quadrupolar $n=2$ harmonic and has significant 
components from higher harmonics \cite{Peters:1963ux,Pierro:2000ej}.  These effects do not, by and large, change the orders of magnitude for our estimates.

In order to estimate the proposed signal strengths, we will need to make sure that parameters of the orbits we examine 
can yield valid results.  In this regard, we need to know the last stable orbit (LSO) for our systems.  According to the results in 
Ref.~\cite{Cutler:1994pb}, for a test particle going around a black hole of mass $M$ in an orbit with eccentricity $e$, the radius of the LSO is given by
\beq
r_{\rm LSO} = \frac{G_N}{c^2} \frac{(6+2e)M}{1+e}, 
\label{rLSO}
\eeq
where $G_N = 6.67 \times 10^{-8}~\text{cm}^3 \text{g}^{-1} \text{s}^{-2}$ is Newton's constant and $c = 3.00 \times 10^{10}$~cm/s is the speed of light.  For a
circular orbit with $e=0$ we get the familiar result $r_{\rm LSO}= 3 \rsch$ and for $e=1$ we find $r_{\rm LSO}=2\rsch$. Hence, as long as we  
choose $r>3 \rsch$, we can assume stable orbits in our analysis.  For simplicity, we will use $r_{\rm LSO}=3 \rsch$ for both 
the NS and ABH cases.  The results of Ref.~\cite{Torok:2014ina} suggest that this would also be a good estimate for the NS case.

Gravitational waves cause oscillations in the local metric as they travel through spacetime.  These oscillations give rise to strain, {\it i.e.} variations in physical 
length scales, the size of whose amplitude we denote by $h$.  Measurement of strain is the basis of gravitational wave detection.  In the following, non-relativistic speeds and orbits large compared to radii of the compact stellar objects are assumed.  The simple formalism that we will use suffices to get reasonable order of magnitude estimates.  
See {\it e.g.} Ref.~\cite{GW} for an accessible presentation and Ref.~\cite{MTW} for a detailed exposition to the relevant subjects.   
  
For a binary system, 
with component masses $M_1$ and $M_2$, in a circular 
orbit of size $r$ at a distance of $d$ from the observer, we have 
\beq
h= \frac{4 G_N^2}{c^4 }\frac{M_1 M_2}{r \, d}\,.
\label{h}
\eeq 
  
The frequency of the 
corresponding gravitational waves are then given by 
\beq
f =\frac{1}{\pi}\left[\frac{G_N(M_1 + M_2)}{r^3}\right]^{1/2}.
\label{f}
\eeq
The radiation of gravitational waves by the binary system causes the decay of its orbital radius 
$r$ to a smaller radius $r_f$ after a time \cite{Peters:1964zz}
\beq
\Delta t_f(r) = \frac{5\, c^5}{256\, G_N^3}\left[\frac{r^4 - r_f^4}{M_1 M_2 (M_1+M_2)}\right]\,.
\label{deltat}
\eeq 
Of particular interest is the time $\Delta t_{\rm LSO}$, which we obtain from \eq{deltat},  required for the system to evolve to the LSO at $r_f=r_{\rm LSO}$.

For concreteness, we will consider $f_* = 150$~Hz as a typical value 
where aLIGO/AdV reach for gravitational waves is optimal.  Our estimates do not sensitively depend 
on the exact value of $f_*$ near our reference value.  Using our reference values in \eq{masses},  \eq{f} yields the radius $r_*$ corresponding to $f_*$
\beq
r_* \approx 97~\text{km}\; \text{(NS)}\quad \text{and} \quad r_* \approx 182~\text{km}\; \text{(ABH)}.
\label{r*}
\eeq
Note that for the ``D\&G" binaries of interest 
here, we have $\mpbh\ll \msol$ and hence the frequency $f_*$ of the waves is independent of 
$\mpbh$, to a very good approximation.  We see that for the above choice of parameters, $r_*$ is well above the radius of the 
NS, about 10~km, and the implied value of $r_{\rm LSO}$ from \eq{rLSO}.

\begin{figure}[t]
\includegraphics[width=0.48\textwidth,clip]{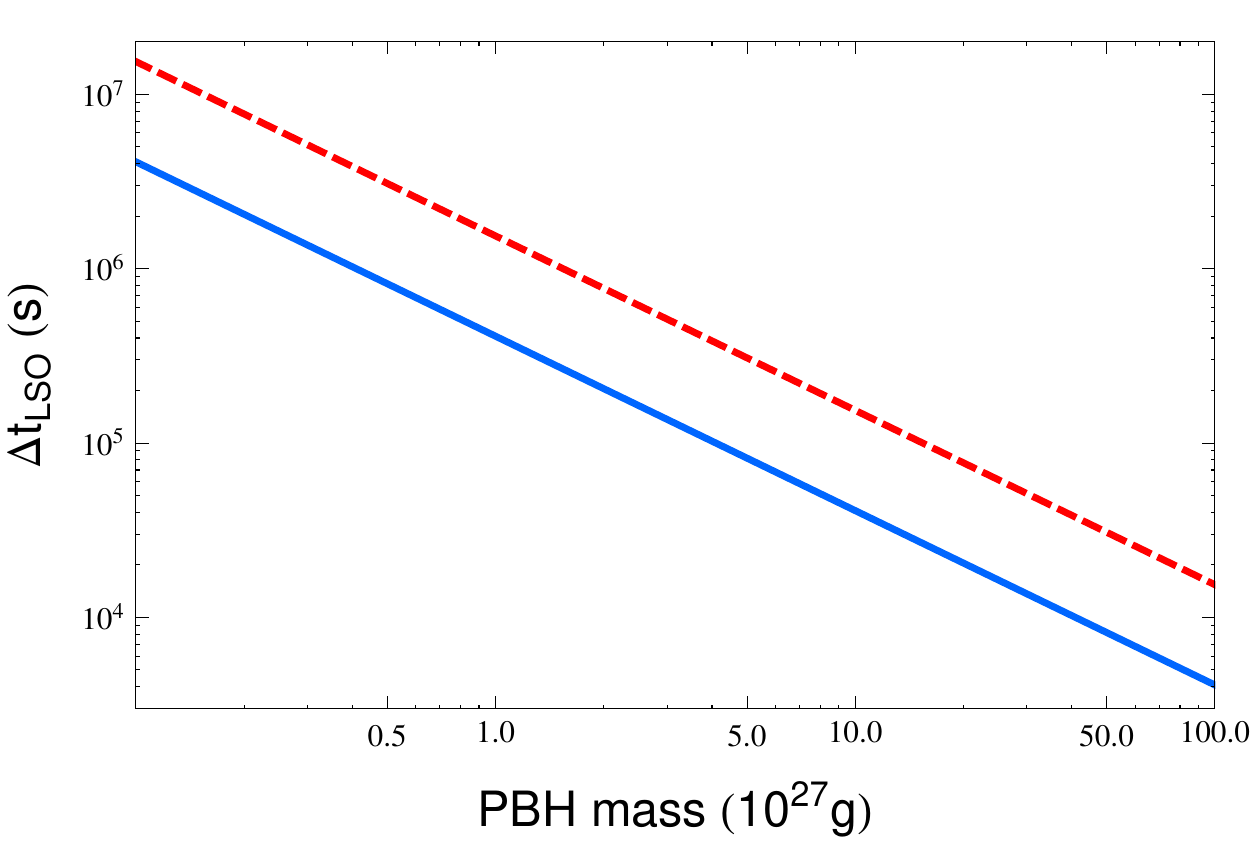}
\caption{Time, in seconds, required for the binary with $\mns = 1.5 \msol$ (dahed) and $\mabh=10\msol$ (solid) 
to evolve from an orbit where it emits gravitational waves at $f_*=150$~Hz to the last stable orbit given by $r=3 \rsch$ (assumed for both the NS and ABH cases, 
using the corresponding mass).}
\label{fig:tm} 
\end{figure}

The decay time $\Delta t_{\rm LSO}$ versus $\mpbh$ is plotted in Fig.\ref{fig:tm}, for $\mns=1.5\msol$, 
$\mabh = 10 \msol$, and $f_*=150$~Hz.   
We see that  $4\times 10^3~\text{s} \lsim \Delta t_{\rm LSO} \lsim 10^{7}~\text{s}$.  We will choose the``observation time" 
\beq
t_{\rm obs} =  \Delta t_{\rm LSO}, 
\label{tobs}
\eeq
which we will assume over the parameter space of our analysis. 

\begin{figure}[t]
\includegraphics[width=0.5\textwidth,clip]{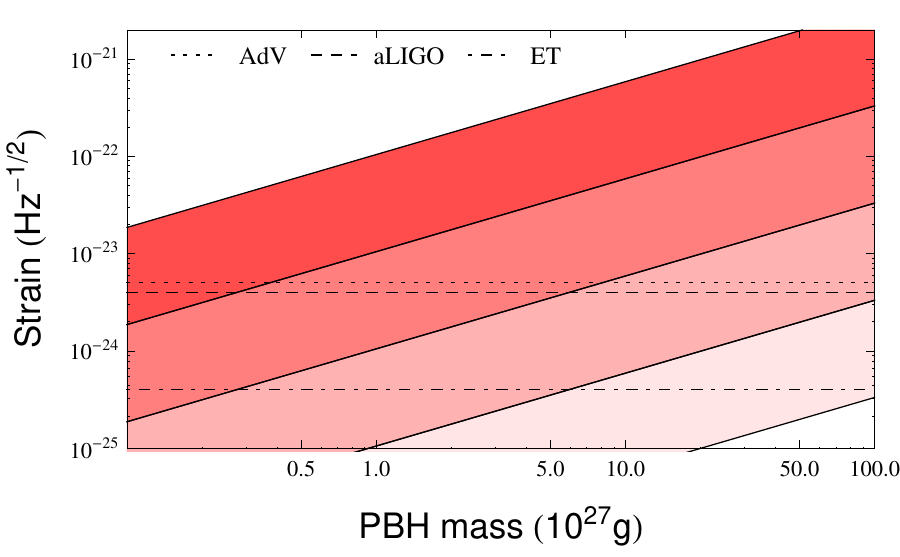}
\caption{Gravitational wave strain signal, in $\text{Hz}^{-1/2}$, for a PBH-NS binary system, as a function of $\mpbh$, with $\mns=1.5\msol$ and $r=r_*$, corresponding to a frequency of $f_*=150$~Hz. Different shades of red from darker to lighter correspond to the distance $d$ intervals, $(5,50)$~kpc, $(50, 500)$~kpc, $(0.5, 5)$~Mpc, and $(5,50)$~Mpc.  
An observation time of $t_{\rm obs}=N\,t_{\rm coh}$ has been assumed, using \eq{tobs} and $t_{\rm coh}=2000$~s.   The horizontal dotted, dashed, and dot-dashed lines represent the expected final design sensitivities for AdV, aLIGO, and ET, respectively.}
\label{fig:hmNS} 
\end{figure}

\begin{figure}[t]
\includegraphics[width=0.5\textwidth,clip]{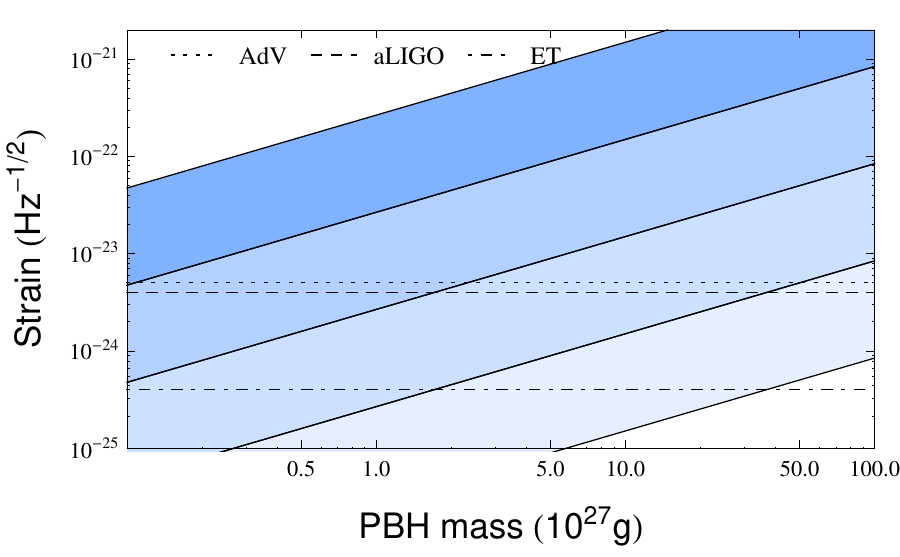}
\caption{Gravitational wave strain signal, in $\text{Hz}^{-1/2}$, for a PBH-ABH binary system, as a function of $\mpbh$, with $\mabh=10\msol$ and $r=r_*$, corresponding to a frequency of $f_*=150$~Hz. Different shades of blue from darker to lighter correspond to the distance $d$ intervals, $(5,50)$~kpc, $(50, 500)$~kpc, $(0.5, 5)$~Mpc, and $(5,50)$~Mpc.  
An observation time of $t_{\rm obs}=N\,t_{\rm coh}$ has been assumed, using \eq{tobs} and $t_{\rm coh}=2000$~s.  
The horizontal dotted, dashed, and dot-dashed lines represent the expected final design sensitivities for AdV, aLIGO, and ET, respectively.}
\label{fig:hmABH} 
\end{figure}

In Fig.\ref{fig:hmNS}, we have plotted the expected size of the strain signal 
$h N^{-1/4}\sqrt{t_{\rm obs}}$, with $t_{\rm obs}=N\,t_{\rm coh}$, 
versus $\mpbh$ for $\mns= 1.5 \msol$ and 
distance from Earth $ 5~\text{kpc} \leq d \leq 50\text{Mpc}$.  Here, $t_{\rm coh}$ 
is the time scale over which the signal can be coherently observed.
The value of $r_*$ has been chosen from \eq{r*} corresponding to the NS case.  
The horizontal dotted, dashed, and dot-dashed lines mark the projected AdV, aLIGO, and the proposed Einstein Telescope (ET) \cite{Punturo:2010zza} 
sensitivities at $f=f_*$, in $1/\sqrt{\text{Hz}}$, of approximately $5\times 10^{-24}$, $4\times 10^{-24}$  \cite{Aasi:2013wya}, and $4\times 10^{-25}$ \cite{Hild:2010id}, respectively.  We have used  $t_{\rm coh}=2000$~s (see for example Ref.~\cite{Aasi:2013lva}) in obtaining the results in Fig.\ref{fig:hmNS}.  We see that for most of the 
range of $\mpbh$ considered here, the entire Milky Way ($d\lsim 50$~kpc) is within the reach of aLIGO/AdV.

We note that the rate of the frequency increase for the systems we examine is intrinsically quite slow, and one could also focus the search on $\ord{2000}$ 
known ``pulsars" in our Galaxy whose optical signals determine their positions in the sky.  This feature allows one to account for signal modulation due to the motion of 
the observer with respect to the barycenter of the Solar System, which may lead to $t_{\rm obs}=t_{\rm coh}$, enhancing the reach for Galactic NS-PBH systems.  

The values of $h N^{-1/4}\sqrt{t_{\rm obs}}$  versus $\mpbh$ are given in Fig.\ref{fig:hmABH}, 
for the ABH case is Eqs.(\ref{masses}) and (\ref{r*}), where we have again
assumed $t_{\rm coh}=2000$~s.  Our results in Fig.\ref{fig:hmABH}
suggest that for $\mpbh \approx 10^{29}$~g, aLIGO/AdV can be sensitive to the 
gravitational wave signals of a PBH-ABH binary out to distances of $\ord{10}$~Mpc, while 
ET can probe $d\lsim 50$~Mpc, beyond our Local Supercluster.  

Note that our signal will not be mistaken for that of a small planet or asteroid captured 
around an NS or ABH.  This is because our gravitational wave signals are obtained for $r_*\sim 100$~km.  This should be compared to the much larger 
radius of the Earth $R_\oplus\sim 6000$~km, whose mass $M_\oplus\sim 6\times 10^{27}$~g is in the $\mpbh$ range of our proposal.   In any event, 
a compact star will tidally destroy a terrestrial scale rocky object, well before reaching an orbit comparable to its size.

In conclusion, we illustrated, as a proof of principle, that if a primordial black hole of mass $\sim 10^{26}$--$10^{29}$~g is captured by a neutron star 
or an astrophysical black hole in our galactic neighborhood, gravitational wave signals of  their ``D\&G" confrontation could be detected by aLIGO/AdV or the proposed Einstein Telescope.  Current constraints allow these primordial black holes to constitute a significant fraction of cosmic dark matter.  Although the signals we consider might be rare, their discovery could shed light on early Universe phase transitions in the visible and hidden sectors relevant to weak scale phenomena.  As such, we may also expect that our signals may be accompanied by 
discovery of new states $\sim 10-100$~GeV and also long wavelength primordial gravitational waves from the phase transition era.  Therefore, we believe that searching for these signals in the existing and future data is well motivated.  

The observation of gravitational waves by LIGO has opened an exciting new front in the exploration of the Cosmos.  We hope that our work would further expand the range of questions  that could potentially be examined at this front.

\acknowledgments

We thank Scott Hughes for very helpful comments and constructive criticism regarding our proposal and Tongyan Lin for useful discussions.
This work is supported by the United States Department of Energy under Grant Contract DE-SC0012704.


\appendix

\section{Appendix}

Here, we provide an order of magnitude estimate for the rate of D\&G binary signal.  As discussed before, the binaries may form either in the process of star formation, via the capture of a PBH by a massive star whose remnant later forms a binary with the PBH, or through radiative capture.  Here, we focus on the second possibility, and estimate the rate for an 
ABH to capture a PBH through gravitational radiation; the realistic rate may potentially be larger.  Also, there is some contribution from NS-PBH binaries that could add to the expected signal rate.  In any event, given the multitude of contributing factors, the following should be viewed as a rough guide.

Following the discussions in Refs.~\cite{O'Leary:2008xt,Cholis:2016kqi}, let $\eta\equiv \mpbh M/\mtot^2$, where 
$M$ is the mass of the NS or ABH and $\mtot\equiv \mpbh + M$. The maximum impact parameter $b$ that leads to the formation 
of the binary, assuming a relative velocity of $w$, is given by 
\beq
b_{\rm max} = \left(\frac{340 \pi}{3}\right)^{1/7} \frac{\mtot \,\eta^{1/7}}{w^{9/7}} G_N c^{-\frac{5}{7}}.
\label{bmax}
\eeq

We will choose $\mpbh\sim10^{29}$~g, since it offers the farthest reach 
in our range of PBH masses in (\ref{Mpbh}) as seen from Fig.\ref{fig:hmABH}, and set $M=\mabh\sim10 \msol$.  The results of 
Ref.~\cite{Gammaldi:2016uhg} suggest that
within the inner 100~pc of the Milky Way, one could have a DM content of 
$\sim 4\times 10^8\,\msol$, though this quantity has large uncertainties.  Hence, asuming some enhancement of DM density towards smaller radii, we can reasonably assume that the DM mass 
contained within the central 10~pc of the Galaxy is $\sim 10^6 \msol$.  The simulations of 
Ref.~\cite{Antonini:2014spa} also imply that $\sim 10^5$ 
ABHs of mass $10 \msol$ could be contained within the same radius.  Hence, the contribution of 
DM (including a sub-dominant PBH population) and ABHs can be comparable and of order $10^6 \msol$.  Assuming that the total mass within 10~pc of the center of the Galaxy is $\sim \text{few}\times 10^6\msol$, we find that 
$w \sim 30~\text{--}~40$~km/s can be a fair estimate.    

For the above set of parameters, one finds the cross section $\sigma_{\rm 
ABH} \sim \pi b_{\rm max}^2 \sim 10^{12}$~km$^2$.   
Assuming that 
the  PBHs are distributed around the value chosen here, we find a PBH 
number density of $n_{\rm PBH} \sim 10^{-34}$~km$^{-3}$.

We may then estimate the capture rate for D\&G binaries of 
interest, near the core of the Milky Way, as
$R \sim \sigma_{\rm ABH} \, n_{\rm PBH} \, w \, N_{\rm ABH} 
\sim 10^{-8}$~yr$^{-1}$.  Here, $N_{\rm ABH}\sim 10^5$ 
is the number of ABHs within the
inner $\sim 10$~pc of the Galaxy.  Given our results in Fig.\ref{fig:hmABH}, we may assume 
that for the chosen parameters aLIGO/AdV could be sensitive to sources $\sim 10$~Mpc away, which covers most of the Local Supercluster, comprising $\sim 2000$ large galaxies.  Hence, we may roughly set the expected rate for aLIGO/AdV at $\sim \text{few}\times 10^{-5}$~yr$^{-1}$.  This rate could potentially be enhanced if we also include expected signals from 
NS-PBH mergers, as well as binary formation processes besides radiative capture.  Therefore, we can tentatively assume a signal rate $\lsim 10^{-4}$~yr$^{-1}$.


\end{document}